\documentstyle[12pt]{article}
\begin{document}

\def\simgt{\ \raisebox{-.25ex}{$\stackrel{>}{\scriptstyle \sim}$}\ }
 \def\simlt{\ \raisebox{-.25ex}{$\stackrel{<}{\scriptstyle \sim}$}\ }
\def\anti#1{\overline#1}
\def\solar{_\odot}

\begin{titlepage}
\vspace{.3in}
\begin{center}
{\Large{\bf Anthropic considerations in multiple-domain 
theories and the scale of electroweak symmetry breaking}}\\
\vskip 0.3in
{\bf V. Agrawal${}^a$, S.M. Barr${}^a$, John F. Donoghue${}^b$
 and D. Seckel${}^a$}\\[.2in]
a){\it Bartol Research Institute \\
University of Delaware, Newark, DE 19716\\}
b){\it Department of Physics and Astronomy, \\ 
University of Massachusetts,
Amherst, MA 01003 \\}

\end{center}
\vskip 0.4in

\begin{abstract}
One of the puzzles of the Standard Model is why the mass parameter
which determines the scale of the Weak interactions is closer to
the scale of Quantum Chromodymanics (QCD) than to the Grand
Unification or Planck scales. We discuss a novel approach to this
problem which is possible in theories in which different regions of the 
universe can have different values of the the physical parameters. In 
such a situation, we would naturally find ourselves in a region which 
has parameters favorable for life. We explore the whole range
of possible values of the mass parameter in the Higgs 
potential, $\mu^2$, from $+M_P^2$ to $-M_P^2$ and find that there is 
only a narrow window, overlapping the observed value, in which 
life is likely to be possible. The observed value of $\mu^2$
is fairly typical of the values in this range. 
Thus multiple domain theories in which $\mu^2$ varies among domains may give
a promising approach to solving the fine tuning problem and explaining
the closeness of the QCD scale and the Weak scale.

\end{abstract}
\end{titlepage}

In our present theory of physics, there are only three parameters in
the fundamental Lagrangian which are dimensionful. Two of 
these are associated with General Relativity, i.e. the Planck Mass 
$M_P^2 = G_N^{-1} = $($10^{19}$ GeV)$^2$, and the cosmological constant,
which is presently bounded to be $\Lambda \le 10^{-120}M_P^4$. The
third is the mass parameter in the Higgs potential of the Standard
Model, $\mu^2$, which leads to a vacuum expectation value for the
Higgs field $v = \sqrt{{-\mu^2 / \lambda}} = 246$ GeV. 
($\lambda \sim 1$.) The expectation value $v$
is the origin of the masses of all of the quarks, leptons and gauge
bosons. A fourth mass scale does not appear in the Lagrangian, but
enters indirectly as the energy at which the ``running" strong coupling
constant becomes of order unity. This QCD scale is roughly 200 MeV.
Because the QCD coupling varies only logarithmically with the energy,
it is natural that the QCD scale is much smaller than the Planck
Mass. However, the smallness of the cosmological constant and the
Higgs mass parameter are severe problems for our present
understanding.  

The Higgs vacuum expectation value is not only small compared to the
Planck scale, $v \sim 10^{-17} M_P$, but it is also problematic
because it receives large quantum corrections. If the Standard Model
is the appropriate description up to some scale $\Lambda_{SM}$, then
$\mu^2$ receives radiative corrections of order
$\Lambda_{SM}^2$. For the Standard Model to be valid
to high energies ($\Lambda_{SM} >> v$), one requires a highly fortuitous
cancellation of the bare parameter and its radiative corrections in order to
produce a low physical value of $\mu^2$. The puzzling smallness of $\mu^2$ is
often referred to as the ``hierarchy problem'', and the sensitivity to
quantum corrections as the ``fine-tuning problem''[1]. The smallness and
fine-tuning of the cosmological constant is even more dramatic [2].

The problem of the Higgs mass parameter is one of the key issues in
modern particle physics, and has led to the widespread expectation
that new physics beyond the Standard Model must be present at energies
$\Lambda_{SM} \sim 1$ TeV. Prime candidates are supersymmetric
theories [3] or theories without fundamental Higgs fields [4]. 
The search for this new physics is a prime goal of theoretical and
experimental efforts.

However, there is the possibility of an entirely different hypothesis, 
in which one posits certain new cosmological features which would 
naturally imply
``anthropic" [5] constraints on some parameters. In exploring 
theories of inflation,
the possibility has emerged that different domains of the universe
could involve different values of the fundamental parameters. 
In such
theories, typical of chaotic inflation [6], dynamical Higgs-like fields
can get fixed at various vacuum expectation values, defining
low-energy theories with different parameters. Our observed universe
would be entirely within one such domain.
The idea of multiple domains may be more general than chaotic
inflation and may potentially be realizable in other contexts also [7].
With our present limited information, it is not any more scientific to
assume that only one unique domain exists than it is to explore the
possibility of multiple domains. The idea that multiple domains
may exist takes the
Copernican revolution to its ultimate limit --- even our universe may
not be the center of the Universe.

Within such a theory it is an obvious
requirement that out of the ensemble of all domains we could only find
ourselves in domains in which
physical parameters are such as to allow the 
development of life --- we will call these ``viable" domains. 
This may drastically narrow the range of allowed values for the mass
parameters. Weinberg has already used this form of reasoning to argue [7]
that the ``anthropic" need for the clustering of galaxies can only be
possible for cosmological constants which are smaller than a value
which is close to the present bound. In this paper, we argue that
under the assumption that life requires the complex elements to be 
formed in the universe one has a constraint that only
allows values of $\mu^2$ close to the QCD scale and in a
range near that found in our domain. If the multiple-domain
cosmological theories are correct, this limited allowed range would
plausibly provide an explanation for the observed small value of the mass
scale of the Standard Model [8].

In the process, these considerations will also illuminate another
puzzle posed by the Standard Model. Even if a different kind of
mechanism to solve the
fine-tuning problem is found, and a hierarchy of scales is allowed, it 
is puzzling that, out of all the available parameter space, the weak
scale is intertwined with the QCD scale. Quark and lepton masses
(manifestations of the weak scale) appear at values both below and
above the QCD scale, and to describe the physical world we need
important inputs from both weak scale physics and QCD physics.
Within the Standard Model, there is no need for these scales to be
close. As far as we know, even in extended theories there is no known 
explanation for this curious fact. Logically, the fine-tuning problem,
the hierarchy problem and this ``intertwined scales'' problem are all
distinct, although they are all aspects of our need to understand the
scale of weak symmetry breaking.

We consider all values of $\mu^2$ from $-M_P^2$ to $+M_P^2$, under the
condition that all
dimensionless parameters of the Standard Model are held fixed at
the unification or Planck scale.
Many of our arguments could be adapted to situations where more
parameters vary, although without knowing more about the underlying
theory one cannot be sure which parameters should be treated as
variable. Our results
are displayed compactly in Fig. 1, and the rest of this paper is
devoted to explaining this figure. The key ideas are relatively simple
to present, and we provide more details in a 
longer paper [9]. We label the values of parameters found in our
domain by a subscript zero, i.e. $\mu_0^2$ and $v_0$.

The impact of the variable values of $\mu^2$ and $v$ are transmitted
to the structure of the chemical elements largely through the quark
and lepton masses, since these are linearly proportional to $v$, i.e. $m =
m_0(v/v_0)$. The most important of these are the up and down quarks
(with $m_u/m_d = 0.6$, $m_{d0} \sim 7$ MeV) and the electron ($m_{e0} =
0.5 $ MeV). Despite the electromagnetic mass shift which enhances the
proton mass ($(m_p - m_n)_{EM} \sim 1.7$ MeV), the neutron is
heavier than the proton because of the larger down quark mass.
The quark masses also play a role in the nuclear force, most
importantly through the attractive long-range pion-exchange potential
which has a range $r \sim 1/m_\pi$, with the pion mass-squared
roughly linearly proportional to the light quark masses, $m_\pi^2
\propto (m_u+m_d)$.

If we start close to the observed values, we note that smaller values
of $v$ appear to be allowed. As $v$ becomes smaller, the nuclear
binding becomes more effective (see the discussion below) and for
$v$ less than a critical value, which we
we estimate to be about $0.7 v_0$ to $0.85 v_0$, the di-neutron and di-proton
become bound. This has a large impact on the relative abundances of
elements [10], but does not prevent
the existence of complex nuclei. Stellar evolution is greatly affected. 
It is amusing to note that below $v/v_0 =
0.5$ the proton is heavier than the neutron and decays $ p \to n e^+
\nu$. In such a domain there would be no hydrogen, and much of matter
would consist of neutrons. However, deuterium and the complex elements
would still exist and could have enough potential to produce life of
some form. We see no clear reason why domains with $v < v_0$, and even
close to zero, would not be biologically ``viable". 

For values of $v$ larger than $v_0$, the elements will become
increasingly unstable. The first key element to become unbound will be
the deuteron, which is just barely bound in nature. As the nuclear
force becomes shorter range with increasing $v$, we estimate that
deuterium becomes unstable against the strong decay $d \to p+n$ at some
value of $v/v_0$ in the 
range $1.4$ to $2.7$ depending on the model used for the nucleon-nucleon
potential. This presents an obstacle to the formation of the elements, as
both nucleosynthesis in the early universe and in the burning of stars
requires a stable deuteron for the initial processes. Beyond this
critical value of $v/v_0$, a domain would likely lack most of the 
elements required for life. However, even if there were a way to form the 
elements, a more severe problem develops at a value of $v/v_0$ around $5$. 
At values larger than this the neutron is heavier than the proton by more 
than the nucleon's binding energy in nuclei, so that even bound neutrons 
would decay to protons. (Of course, as $N$ becomes less than $Z$ in this way, 
the change in the nuclear fermi energies make  $n \longrightarrow p e^- 
\overline{\nu}$ less exothermic, but our understanding of nuclear structure 
indicates that nuclei with $Z \gg N$ are not bound anyway.) Such a domain
would contain only protons, would not form complex nuclei, and would be 
chemically sterile, and therefore probably not viable. This yields our 
first bound 
on $\mu^2$ on the left side of Fig. 1. It is interesting that the existence
of neutrons close enough in mass to the proton to be stable in nuclei
appears to be a requirement for life to exist.

Domains with $v/v_0$ above $5$ and below another critical value near 
$10^3$ would appear as sterile ``proton domains". In domains with
$v/v_0$ above around $10^3$ the only stable baryons would be
$\Delta^{++}$ particles, which, 
being atomically equivalent to helium, would be even more chemically 
inert. This transition to ``$\Delta$ domains" happens when the $d-u$ 
mass difference is 
large enough that the $\Delta^{++}$ (i.e. $uuu$) is lighter than the 
proton ($uud$) despite the QCD hyperfine energy which shifts the 
$\Delta$'s up in mass by about 300 MeV compared to the proton. We have 
estimated the non-relativistic binding energy of 6 ultra-heavy $u$ quarks 
in a single object and find that almost certainly it would fission to two 
$\Delta^{++}$'s. (At the transition point between ``proton domains"
and ``$\Delta$ domains", there is a narrow range of $v/v_0$ where
the electron mass would stabilize both $p$ and $\Delta^{++}$, but even 
this somewhat richer chemical environment seems unlikely to support life 
processes.) 

Where $\mu^2$ has the opposite sign from that in our domain, the Higgs 
potential does not lead to electroweak symmetry breaking; rather, the 
$SU(2)_L$ symmetry is broken by the chiral dynamics of QCD. This leads to 
light $W^{\pm}$ and $Z^0$ gauge bosons ($M_W \sim 50$ MeV). It also leads 
to a tiny value for $v \sim f_\pi^3/\mu^2$, so all the quarks and leptons 
are nearly massless. This leads to domains which are very different 
from our own, hard to analyze definitively, but with several features that 
appear to disfavor the possibility of life. 

All energy scales in chemistry are set by the electron mass, which for 
$\mu^2 > |\mu_0^2|$ would be smaller by more than a factor of a billion. 
Chemical binding energies would therefore be very small. 
It is clear that chemical life cannot emerge until the time, $t_{chem}$ 
when the temperature of the universe cools below typical biochemical 
reaction energies; otherwise (to put it picturesquely) life would be 
fried by the primordial cosmic background radiation. For 
electron-dominated chemistry we estimate 
\begin{equation}
t_{chem} \sim  10^{23} {\rm yrs}
\; 
\left( \frac{\mu^2}{|\mu_0^2|} \right)^{\frac{3}{2}}.
\end{equation}
This timescale could be reduced by a factor of up to 50 if 
the valence electrons were replaced by muons and/or tau leptons, which 
are effectively stable due to their small mass. In any event 
$t_{chem}$ is a long time and several factors relevant to the development
of chemical life would be altered. For example, if life is to evolve it 
must do so before all the baryons decay, or before all stars reach the 
end of their evolutionary paths.

It is likely that baryons can decay. The unification of gauge 
couplings [11] suggests the existence of gauge bosons of mass $10^{16}$ GeV 
whose exchange leads to violation of baryon number. Even without this, it 
is plausible that Planck scale physics leads to baryon decay. We 
therefore parameterize the baryon decay rate as $\Gamma_B = m_p^5/M^4$, 
where $M$ is assumed to lie between $10^{16}$ GeV and $10^{19}$ GeV. In 
comparing $t_{chem}$ to $\Gamma_B$ we must include the 
thermalized energy from the decaying baryons, which was left out of Eq. 1.
The temperature at the epoch of baryon decay 
will be $T_{rad} \sim (\Gamma_B M_P)^{1/2}$. If $T_{rad}$ is greater than 
some fraction (which in our universe is of order $10^{-3}$) of the energy
binding leptons to atoms, then life based on chemistry will be impossible. 
This constraint rules out the larger positive values of $\mu^2$ as not
being biologically viable, 
as shown for $M=10^{16}$ GeV and electron chemistry in our Figure. 
This constraint could be much stronger if baryon decay involves 
the exchange of fermions (as in many supersymmetric models) or weak
interaction processes known as sphalerons [12], both of which are suppressed 
in our world but may be allowed in a world with ultralight quarks and 
and QCD-mass-scale weak bosons.

Even if baryons exist, it is not clear if or how they would form nuclei 
appropriate for chemical life to evolve. Since all the quarks are light: 
a) the ground-state baryons will contain 27 members, including the  
neutron and proton, and b) there will be a host of neutral mesons with 
masses less than a KeV (for $\mu^2>|\mu_0^2|$). Nuclear forces will be 
long-range, although short-range repulsive forces would still lead to a 
saturation of nuclear density. The large number of nucleon species will 
produce lower fermi levels in nuclei. Since weak forces have a range of 
several fermis, their contribution to the electrostatic energy must be 
included. For intermediate size nuclei (a few $<A<$ a few hundred) we 
find that $Z \sim A/4$. The finite range of the weak force means that
in very large nuclei $Z$ and $N$ will adjust to minimize electrostatic
rather than electroweak energy, and thus $Z$ will be much smaller than $A$.
Given the uncertainties, it is unclear whether or not there is a maximum 
nuclear size beyond which spontaneous fission occurs. 

The long range of mesonic nuclear forces suggests that nucleosynthesis 
will proceed rapidly. However, in a thermal bath the effective mass of 
the mesons will be significant, and the range of nuclear forces will be 
reduced. It is unclear whether electrostatic coulomb and weak potentials 
will provide an effective barrier to nuclear reactions in a plasma. If 
they do, then primordial nucleosynthesis will halt at modest charges and 
nuclear sizes. There will be ample fuel for stars and a plausible 
elemental mix for life. If not, then primordial nucleosynthesis will run 
away either to the equivalent of trans-iron elements or to super-heavy 
nuclei with very low ratios of charge to mass. It is questionable if 
either of the last two scenarios would lead to biologically viable
domains. 

Even if nucleosynthesis produces an appropriate mix of 
elements, there is a question of stellar evolution and finding an 
environment and energy source for life to develop. With extremely light 
leptons, objects with mass less than a solar mass ($M<M\solar$) will 
condense to planets supported by non-relativistic degenerate leptons. As 
larger objects cool, the leptons become relativistic before they become 
degenerate, and so such objects will condense to stars and burn nuclear 
fuel. The cooling time during the pre-ignition phases of stellar 
evolution will be dominated by photon diffusion at a time when the 
internal temperature is comparable to the electron mass (which maximizes 
the compton cross-section). We estimate $t_{cool} \approx 10^{17} 
\mu^2/\|\mu_0^2|$ yr. This is less than $t_{chem}$, but not by so much 
that stars may not be important as energy sources for life.

If electrostatic coulomb barriers are effective in a plasma of 
charged leptons and neutral mesons, thermonuclear reactions will support 
the star at temperatures of $1-10$ KeV. Because of the ultra-light charged 
leptons, radiative opacities will be large. Therefore, given the small
$W^{\pm}$ and $Z^0$ masses, such an object will cool by neutrino 
pair emission. We estimate nuclear burning lifetimes for $M\sim M\solar$ 
of roughly a year, and much less for larger stars. This is very much less 
than $t_{chem}$. 

Thus, within this crude treatment of stellar evolution, stars are 
expected to form slowly, and then burn nuclear fuel very quickly. But
both timescales appear to be too small for there to be stars left when
the tempretaure of the universe will allow biochemistry. However, it 
is possible that other sources of energy may be available, eg, 
gravitational energy of stars collapsing to the main sequence, ``geothermal" 
energy, energy from radioactive decay, etc. It is therefore plausible, 
but by no means certain, that elemental and stellar evolutionary 
considerations exclude life in $\mu^2 > 0$ domains in the remaining area 
of Figure 1.

In conclusion, in a universe which has a domain structure, and in
which some parameters have different values in different domains,
life may only be able to develop in some domains and not others. 
If this is the case, we would expect that the parameters 
of our domain should be typical of the ``viable" range. We have found 
that within the overall structure of the Standard Model there is a relatively 
small acceptable viable range for the Higgs parameter $\mu^2$. 
It seems that $\mu^2$ must be negative and of absolute magnitude
close to what is observed. A multiple domain scenario in which
$\mu^2$ varied could alleviate the fine-tuning
problem. In an ensemble of different domains, the Higgs
mass parameter will occasionally fall into the viable
range without having to be fine-tuned in general.

If $\mu^2$ is positive, then lepton masses are extremely small, as
$v$ is then set by QCD chiral symmetry breaking. Therefore, biochemical 
energies are also small, and the universe may be so old before it has
cooled sufficiently to allow biochemical life that baryons have all
decayed away, or stars have ceased to form and burn.

If  $\mu^2$ negative, as in our domain, it seems that the whole range 
of values for $v$ from $M_P$ down to about $5$ (or perhaps even down 
to $1.4$) times the value in our domain can be excluded. Any domain
which had a value of $v$ in most of 
that range (down to about $10^3 v_0$) would contain only sterile, 
helium-like atoms whose nuclei were $\Delta^{++}$. There would be essentially
no reactions either chemical or nuclear. For the lower part of the excluded 
range, there would be virtually no nuclei other than protons, and the $pp$ 
and $pn$ processes that are needed for nucleosynthesis would be endothermic 
as the deuteron would not be stable. 

Thus we see that the natural viability requirement present within
multiple domain theories provides a plausible approach to the
fine-tuning problem, the hierarchy problem and the intertwined scales
problem, as well as possibly the cosmological constant problem[7].

Finally,  let us comment on the ability of these ideas to be tested. 
Negatively, we can say that if the weak scale is governed by
``anthropic" considerations,
there would be no need to invoke supersymmetry or technicolor or 
other structure at the 
weak scale to make the fine-tuning ``natural" [1,13]. 
If no such structure is found, it would be a point in favor of anthropic
explanations; indeed, in that case there would be few if any alternatives.
Positive evidence is harder to come by. Of course, we are not able to 
explore other domains in the universe. However, theories which generate 
multiple domains may be testable by other, more conventional means.
Because the class of theories which lead
to multiple cosmological domains is not yet well understood
theoretically, this will certainly be challenging and will not be
completed in the near future.
However, the community is hoping to be able to test the details of 
inflationary theories through cosmological measurements, and this may
possibly inform us on the correctness of chaotic inflation. 
Likewise, as with any theory of physics beyond the Standard Model, we
will require direct physical experimentation to eventually sort out 
the correct underlying theory. Through standard means we may be able to 
learn if the fundamental theory in fact produces multiple domains, and
whether $\mu^2$ can vary among those domains. If so, then the hypothesis
we propose in this paper automatically becomes relevant. Until such time, 
our conclusion must be modest: the observed value of the weak scale is
reasonably typical of the biologically viable range.

\section*{References}

\begin{enumerate}
\item E. Gildener and S. Weinberg, {\it Phys. Rev.} {\bf D13}, 3333 
(1976); E. Gildener, {\it Phys. Rev.} {\bf D14}, 1667 (1976).
\item For a comprehensive review see S. Weinberg, {\it Rev. Mod. Phys.}
{\bf 61}, 1 (1989).
\item S. Dimopoulos and H. Georgi, {\it Nucl. Phys.} {\bf B193}, 150 
(1981); {\it Phys. Lett.} {\bf 117B}, 287 (1982).
\item S. Weinberg, {\it Phys. Rev.} {\bf D13}, 974 (1976);
S. Dimopoulos and L. Susskind, {\it Nucl. Phys.} {\bf B155}, 237 (1979).
\item B. Carter, in I.A.U. Symposium, Vol 63, ed by M. Longair (Reidel,
Dordrecht, 1974);
J. Barrow and F. Tipler, {\it The Anthropic Cosmological Principle}
(Clarendon Press, Oxford, 1986);
B. J. Carr and M.J. Rees, Nature {\bf 278}, 605 (1979);
A. Vilenkin, Phys. Rev. Lett. {\bf 74}, 846 (1995).
\item A. Linde, {\it Phys. Lett.} {\bf B129}, 177 (1983);
{\it ibid.} {\bf B175}, 395 (1986), {\it ibid.} {\bf B202},
194 (1988), Phys. Scri.,{\bf T15}, 169 (1987).
\item S. Weinberg, {\it Phys. Rev. Lett.} {\bf 59}, 2607 (1987);
S. Weinberg, astro-ph/9610044, to be published in the proceedings of the
conference {\it Critical Dialogues in Cosmology} at Princeton University, 
June 1996; H. Martel, P. Shapiro, and S. Weinberg, astro-ph/9701099.
\item R. Cahn, {\it Rev. Mod. Phys.} {\bf 68}, 951 (1996), makes arguments
similar to some of ours, but not in a multiple domain framework, and
not as an approach the gauge hierarchy problem. There
fermion masses are held fixed, whereas here the Yukawa constants are
held fixed and $\mu^2$ varied.
\item V. Agrawal, S. M. Barr, J. F. Donoghue and D. Seckel,
hep-ph/9707380, to be submitted for publication.
\item J.D. Barrow, {\it Phys. Rev.} {\bf D35}, 1807 (1987), and
references therein.
\item H. Georgi, H. Quinn, and S. Weinberg, {\it Phys. Rev. Lett.}
{\bf 33}, 451 (1974). 
\item V. Kuzmin, V. Rubakov, and M. Shaposhnikov, {\it Phys. Lett.}
{\bf B155}, 36 (1985); {\bf B191}, 171 (1987); P. Arnold and L. McLerran,
{\it Phys. Rev.} {\bf D36}, 581 (1987); {\bf D37}, 1020 (1988).
\item Note however that the anthropic arguments are also compatible
with supersymmetry and may illuminate the intertwined scales problem
and the ranges of the many parameters within this theory.

\end{enumerate}

\section*{Figure Caption}

{\bf Figure 1:} The figure shows a summary of arguments 
that $|\mu^2| \ll M_P$ is necessary for life to develop. $\mu^2$
is the mass parameter of the Higgs field of the Standard Model, and
$v$ is its vacuum expectation value. $\mu^2$ can range from $+M_P^2$
to $-M_P^2$. The abscissa is defined to allow both signs of $\mu^2$
to be shown on the same log plot. For $\mu^2<0$, $v \propto 
(-\mu^2)^{1/2}$, and thus large values of $|\mu^2|$ imply large 
masses for leptons, quarks, and baryons. 
The increasing difference between the light quark masses, $m_d - m_u 
\propto v/v_0$, implies universes with but a single species
of stable nucleus ($p$ or $\Delta^{++}$), which we argue would not allow 
for chemistry rich enough to support life. There is a narrow band where 
both $p$ and $\Delta^{++}$ are stable, but the chemical equivalent of a 
mix of hydrogen and helium is probably also sterile. For $\mu^2>0$, quark 
chiral condensates lead to $v \propto f_\pi^3/\mu^2$ (where $f_{\pi} 
\sim 100$ MeV) and quark and lepton masses become very small. Light lepton 
masses imply that biochemical processes cannot occur until cosmologically 
late times, when baryons may have already decayed. We show a constraint 
for a baryon lifetime estimated from exchange of intermediate GUT scale 
($\approx 10^{16}$ GeV) particles. Even if baryons are stable, formation 
of a biologically acceptable mix of elements or the nature of stellar 
evolution may make development of life improbable. What is left is a rather 
narrow range of $\mu^2 <0$ which includes the physical values in our domain.

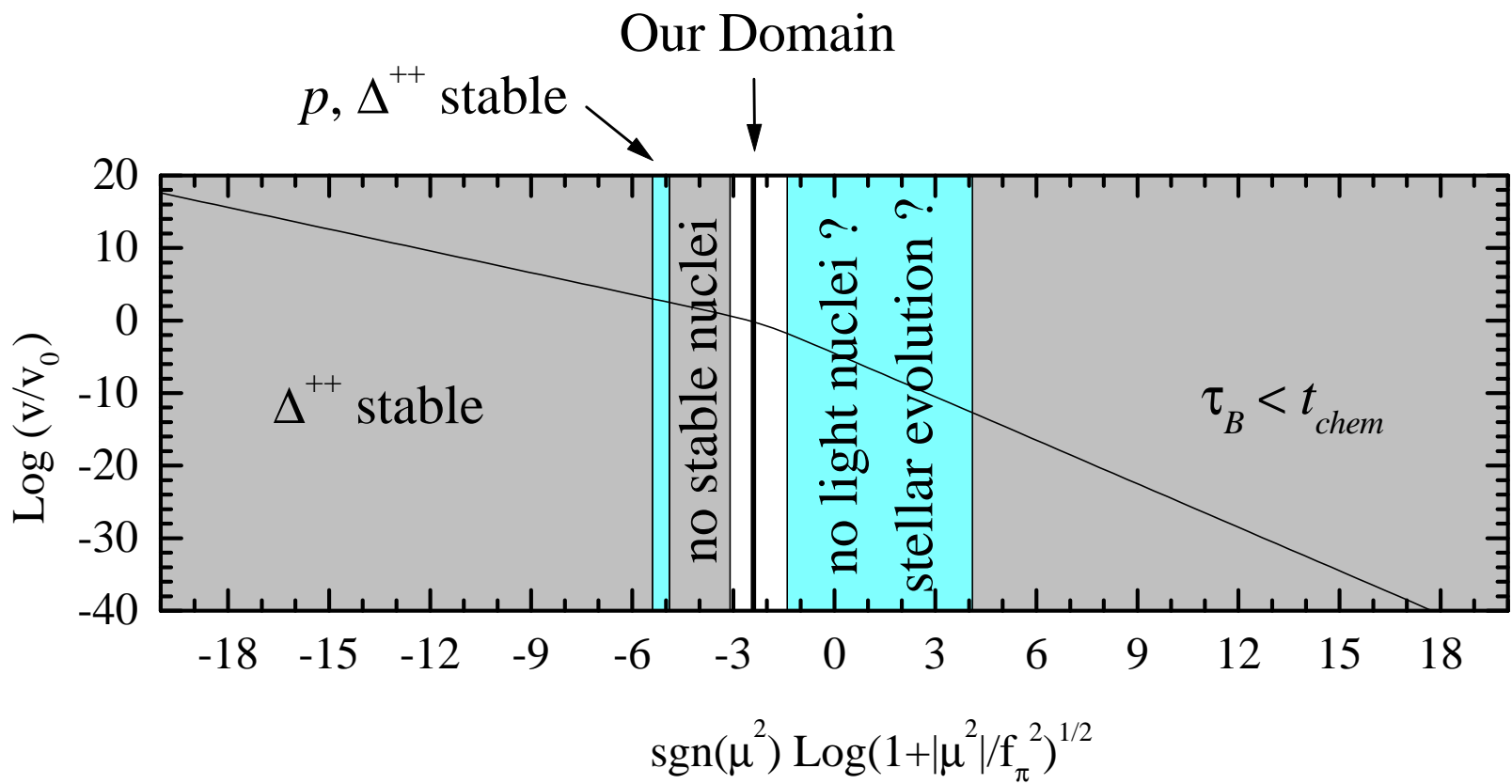
\end{document}